\documentclass[aps,twocolumn,floats,showpacs,superscriptaddress,floatfix]{revtex4}
\usepackage{epsfig}
\def\beginwide{
        \end{multicols} \vspace*{-0.5cm} \noindent
        \rule{3.5in}{.1mm}\rule{.1mm}{5mm} \widetext \medskip }
\def\beginwidetop{
        \end{multicols} \vspace*{-0.5cm} \noindent
        \widetext \medskip }
\def\endwide{
        \hspace*{3.35in}~\rule[-5mm]{.1mm}{5mm}\rule{3.5in}{.1mm}
        \begin{multicols}{2} \vspace*{-1.0cm} \noindent }
\def\endwidebottom{
        \begin{multicols}{2} \vspace*{-1.0cm} \noindent }

\begin{document}
\title{Phase transitions in a disordered system in and out of equilibrium}
\author{Francesca Colaiori}
\affiliation{INFM SMC, Dipartimento di Fisica,
Universit\`a "La Sapienza", P.le A. Moro 2
00185 Roma, Italy}
\author{Mikko J. Alava}
\affiliation{INFM SMC, Dipartimento di Fisica,
Universit\`a "La Sapienza", P.le A. Moro 2
00185 Roma, Italy}
\affiliation{Helsinki University of Technology, Laboratory of Physics,  
HUT-02105 Finland}
\author{Gianfranco Durin}
\author{Alessandro Magni}
\affiliation{Istituto Elettrotecnico Nazionale Galileo Ferraris, 
strada delle Cacce 91, I-10135 Torino, Italy}
\author{Stefano Zapperi}
\affiliation{INFM SMC, Dipartimento di Fisica,
Universit\`a "La Sapienza", P.le A. Moro 2
00185 Roma, Italy}
\begin{abstract}
The equilibrium and non--equilibrium disorder induced phase
transitions are compared in the random-field Ising model (RFIM).  We
identify in the demagnetized state (DS) the correct non-equilibrium
hysteretic counterpart of the $T=0$ ground state (GS), and present
evidence of universality.
Numerical simulations in $d=3$ indicate that exponents and scaling functions
coincide, while the location of the critical point differs, as
corroborated by exact results for the Bethe lattice.  These results
are of relevance for optimization, and for the generic question of
universality in the presence of disorder.
\end{abstract}
\maketitle

Similarities and differences between equilibrium and non--equilibrium 
states in disordered systems have been widely studied both 
for their conceptual importance and because the presence of randomness 
often provides prototypical examples of complex optimization problems 
\cite{ALA-01}. 
There are also many applications in the physics of materials,
where this dichotomy is met, together often with concepts such as aging and
glassiness. The central issue is to understand
whether the equilibrium properties of disordered systems provide a
faithful representation of the non--equilibrium states in which the
system is likely to be found in practice. In optimization terms, 
the question is what is the relation between an approximate solution 
and the optimal one.

A disordered system can be non-trivial even at zero temperature due to
the presence of a complex energy landscape. The properties of the
ground-state (GS) are often difficult to determine analytically, and
numerical evaluation becomes computationally prohibitive for large
systems in particular and for some problems like spin glasses, in
general.  Non-equilibrium dynamics brings the system to the nearest
metastable state and then noise or an applied field is needed to allow
it to explore the energy landscape. Typical optimization methods are
constructed by providing a suitable perturbation scheme on the states
of the system. Recently hysteretic optimization was proposed
\cite{ZAR-02} as an alternative to methods that use noise, such as
simulated annealing \cite{KIR-83}.  Its basis is an analogy to a
ferromagnetic demagnetization procedure: an external oscillating field
with decreasing amplitude and low frequency is applied to the system,
yielding at zero field the demagnetized state (DS), 
which is used as a reference state for material 
characterization \cite{Bertotti}.

The ferromagnetic random field Ising model (RFIM) has been extensively
studied in literature as a paradigmatic example of disordered system
\cite{NAT-00}, whose equilibrium and non-equilibrium properties are still
tractable, though far from trivial.  The RFIM is one of the simplest systems,
where the crucial interplay between quenched disorder and exchange
interaction gives rise in high enough dimensions to a disorder induced
phase transition. This also affects the dynamics, where a
non-equilibrium phase transition exists.

The equilibrium properties of the RFIM are governed by the $T=0$ 
scaling \cite{NAT-00} even at high temperatures. 
GS calculations have elucidated the properties of the phase
diagram: In $d=1$ the RFIM is trivially paramagnetic. In $d=2$ there
is no phase transition but an effective ferromagnetic regime for small
systems. In $d\geq 3$ the GS displays an equilibrium phase transition
induced by the disorder from a low disorder regime where the system is
ferromagnetic (FM) to a strong disorder one, where the system is
paramagnetic (PM) \cite{ALA-01,NAT-00}. The equilibrium critical
exponents for random field magnets have been measured experimentally in 
Fe$_{0.93}$Zn$_{0.07}$F$_2$ \cite{SLA-99,YE-02}.

Likewise, the non--equilibrium properties of the RFIM have been
studied by extensive numerical simulations and renormalization group
calculations \cite{SET-93,DAH-96,PER-99}.  A disorder--induced
transition is observed in the hysteretic behavior for $d \geq 3$: At
low disorder the loop has a macroscopic jump in the magnetization,
which disappears at a critical value of the disorder, above which the
loop is smooth on a macroscopic scale.  Numerical simulations
\cite{PER-99} and renormalization group \cite{DAH-96} have been used
to estimate the critical exponents in various dimensions. A disorder
induced non-equilibrium phase transition in the hysteresis loop has
been studied experimentally in Co-CoO films \cite{BER-00} and Cu-Al-Mn
alloys \cite{MAR-03}.

The relation between this non--equilibrium transition and the PM to FM
one in the (equilibrium) GS has been debated in the past.  Based on
the similarity in the numerical values of the exponents and on
mean--field equations, Maritan {\em et al.} \cite{MAR-94} argued that
the two transitions should be universal.  Sethna {\em et al.}
\cite{SET-94} refuted this due to the different natures of the two
cases: the transition in the GS occurs for a zero external field,
while the transition in the hysteresis loop occurs at the coercive
field.  More recently, the question of the universality of the exponents,
with respect to the shape of the disorder distribution, was discussed
in $d=3$ simulations, mean-field theory, and on the Bethe lattice 
\cite{SOU-97,DUX-01,DOB-02}.

In this letter, we compare the equilibrium and non-equilibrium phase
transitions in the RFIM, with evidence for universality.  The key
issue is the identification of a reference non-equilibrium state,
instead of focusing on the jump in the saturation loop. In the low
disorder phase, a discontinuous hysteresis loop corresponds to a
region of the field--magnetization plane not accessible by any field
history \cite{DAN-02,COL-02}. In this regime it is not possible to
demagnetize completely by applying a slowly varying AC field. Thus one
studies the DS as the state of lower (remanent) magnetization
resulting from the demagnetization procedure. This state is uniquely
defined, in the quasistatic limit, for any given realization of the
random fields. It has two non-equilibrium phases: FM when the main
loop has a jump, and PM otherwise. The remanent magnetization becomes
the order parameter of the transition. Notice that the DS is defined
at $H=0$ and is therefore the natural non--equilibrium counterpart of
the GS. This responds to the objection raised in Ref.~\cite{SET-94}
against the possible existence of universality.

We evaluate in $d=3$ the finite--size scaling functions both for the
equilibrium and non--equilibrium phase transition.  By rescaling the
disorder around the appropriate (distinct) critical values, the
scaling functions can be collapsed onto the same curve using the same
exponents values. However, the location of the critical point differs:
the transition in the DS occurs at a lower disorder value. Thus there
is an intermediate region where the GS is ferromagnetic but the DS is
paramagnetic.  To further corroborate our findings, we analyze the
RFIM on the Bethe lattice: we compute the GS magnetization and compare
it with the remanent magnetization of the DS \cite{COL-02}.  While the
exponents are the same in the two cases (coinciding with mean--field
results), the Bethe lattice reproduces the ordering of the critical
points in $d=3$. Additional evidence for universality is obtained by 
comparing the order parameter distribution function at the critical point 
for finite systems.

In the RFIM, a spin $s_i = \pm 1$ is assigned to each site $i=1,...N$ 
(here of a cubic lattice in $d=3$ or a Bethe one
with coordination $z=4$). The spins are coupled to their
nearest--neighbors spins by a ferromagnetic interaction of strength
$J$ and to the external field $H$. In addition, to each site 
is associated a random field $h_i$ taken from a Gaussian
probability density $\rho(h)=\exp(-h^2/2R^2)/\sqrt{2\pi}R$, with
width $R$, which measures the strength of the disorder. The
Hamiltonian thus reads
\begin{equation} 
{\cal H} = -\sum_{\langle i,j \rangle}Js_i s_j -\sum_i(H+h_i)s_i \,,
\label{eq:rfim}
\end{equation}
where $\sum_{\langle i,j \rangle}$ is restricted to nearest-neighbors pairs.

The RFIM GS is numerically solvable in a polynomial CPU--time with
exact combinatorial algorithms. We find the GS via 
the min--cut/max--flow problem of combinatorial optimization,
and use the so-called push--relabel variant of the preflow algorithm
\cite{eira}. 
Such methods, properly 
implemented, are in general slightly non--linear in their performance
as a function of the number of spins \cite{ALA-01}. 

For the out of equilibrium case, we consider a simple relaxation
dynamics obtained in the limit $T \rightarrow 0$ of the Glauber
dynamics \cite{SET-93,DAH-96,PER-99}: At each time step the spins
align with the local effective field
\begin{equation}
s_i = \mbox{sign}(J\sum_j s_j  + h_i +H)
\end{equation}
until a metastable state is reached. To construct the hysteresis loop,
the system is started from a state with all the spins down $s_i=-1$ and
then $H$ is ramped slowly from $H \to -\infty$ to $H \to \infty$. The
limit of $dH/dt \to 0$ is taken after the limit $T \to 0$. 
In practice, this can be conveniently obtained by precise increases
of the field, to always flip the first unstable
spin. To reach the DS, the external field is changed through a nested
succession $H =H_0 \to H_1 \to H_2 \to ..... H_n...\to 0$, with
$H_{2n}>-H_{2n+1}>H_{2n+2}>0$, and $dH\equiv H_{2n}-H_{2n+2}\to 0$.  
This provides a perfect demagnetization with a uniquely defined DS 
($dH/dt\to 0$). It being quite expensive computationally, we instead
perform an approximate demagnetization using an
algorithm discussed in Ref.~\cite{DAN-02} with $dH =10^{-3}$. 
We verified that the states have negligible differences
with perfectly demagnetized ones.

The RFIM critical exponents
characterizing the disorder induced transition can be defined as
usual: The magnetization $M\equiv \langle |m| \rangle$, with
$m\equiv  \sum_i s_i/N$, scales close to the
transition point as $M=Ar^\beta$,
where $r\equiv (R-R_c)/R_c$ is the reduced order parameter
 and $A$ is a non-universal constant.  The correlation length
exponent $\xi=(Br)^{-\nu}$, where $B$ is another non-universal
constant, rules the finite--size scaling of the model
\begin{equation}
M=A L^{-\beta/\nu}f\left(B L^{1/\nu}(R-R_c)/R_c\right).
\label{eq:fss}
\end{equation}
GS simulations in $d=3$ for Gaussian disorder, yield
$1/\nu^{(GS)}\simeq 0.73$, $\beta^{(GS)}\simeq 0.02$ and
$R_c^{(GS)}\simeq 2.28$ \cite{OGI-86,HAR-99,HAR-01,MID-02}.

The demagnetization process has been exactly solved in $d=1$
\cite{DAN-02} and on the Bethe lattice (\cite{COL-02}, see also
\cite{SHU-01}). Numerical simulations in $d=3$ indicate that the DS
displays the same critical point as the saturation loop \cite{CAR-03}.
The transition point has been obtained numerically in $d=3$ as
$R_c^{(DS)} \simeq 2.16$ \cite{PER-99} and the critical exponents have
been measured.  E.g.Ref.~\cite{CAR-03} reports $\beta_{(DS)}=0.04\pm
0.02$ and $1/\nu_{(DS)}=0.71\pm 0.1$.

\begin{figure}[ht]
\centerline{\psfig{file=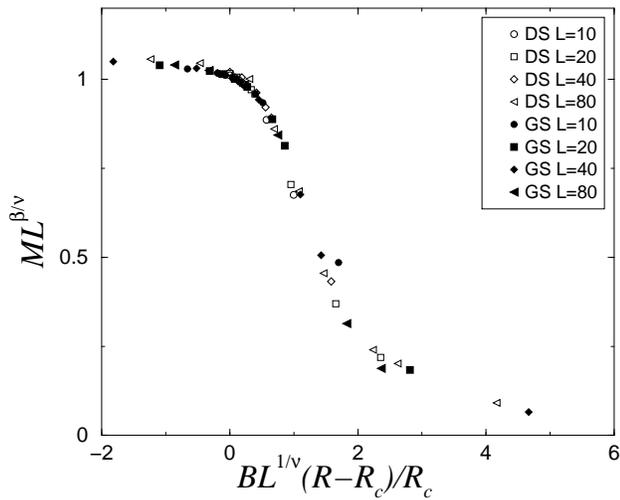,width=8cm,clip=!}}
\caption{Numerical results in $d=£$: The magnetization can be collapsed using  
$R_c=2.28$ (GS) and $R_c=2.16$ (DS), $1/\nu=0.73$ and $\beta=0.03$. 
The scaling curve is the same for DS and GS indicating universal behavior. 
The values for the ratios of the non-universal constants are 
$A_{DS}/A_{GS}= 1$ and $B_{DS}/B_{GS} = 0.68$.}
\label{figcollapse}
\end{figure}

The numerical simulations for the GS and DS are done for the same
disorder realizations for the both cases, for cubic lattices of linear
sizes $L=10,20,40,80$. The results are averaged over several
realizations of the quenched random fields.  In both cases, we compute
the average magnetization as a function of the disorder width.  In
Fig.~\ref{figcollapse} we collapse the two sets of data into a single
curve, using two different values for $R_c$ (i.e. $R_c^{(GS)}=2.28$
and $R_c^{(DS)}=2.16$) but the same values for the exponents
(i.e. $1/\nu=0.73$ and $\beta=0.03$). The best value for the ratio of
the non-universal constant is found to be $A_{DS}/A_{GS}\simeq 1$ and
$B_{DS}/B_{GS} = 0.68 \pm 0.02$.

%fig1

\begin{figure}[h]
\centerline{\psfig{file=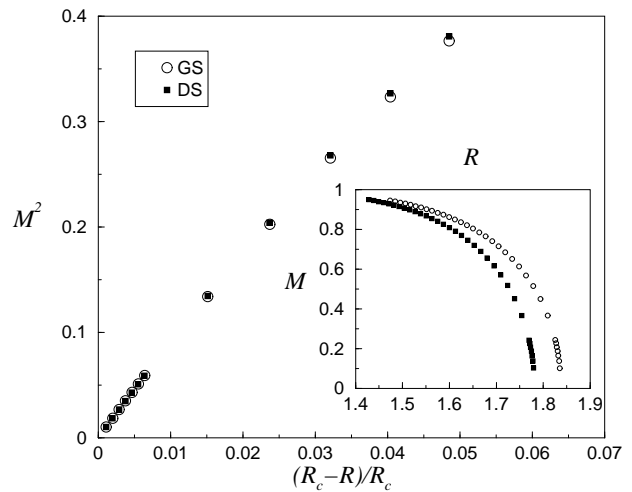,width=8cm}}
\caption{The magnetization of the GS and the DS computed exactly on
the Bethe lattice with $z=4$ in the thermodynamic limit, showing the
ordering of the critical point (see inset).  When the data are plotted
against the reduced parameter $(R_c-R)/R_c$ the curves
superimpose. The result implies that for the Bethe lattice
$A_{GS}=A_{DS}$.}
\label{figbethe}
\end{figure}

To provide another viewpoint and corroborate our claims, we compare
the GS and DS on the Bethe lattice where analytical expressions can be
found exactly. The RFIM displays also on the Bethe lattice, for a
large enough coordination number $z$, both an equilibrium and a
non--equilibrium disorder induced phase transition \cite{COL-02}.  To
compare the GS and the DS around the respective transitions, we take
directly the thermodynamic limit, using for the DS the results of
Ref.~\cite{COL-02}. We have obtained the GS magnetization following
Refs.  \cite{BRU-84,SWI-94} as
\begin{equation}
M=\int_{-\infty}^{\infty} dh \rho(h) \int_{-\infty}^{\infty} 
\prod_{k=1}^{z} dx_k W_{\infty}(x_k) \langle s_0 \rangle.
\label{magb}
\end{equation} 
Here $W_{\infty}(x)$ is fixed-point probability distribution for the
quantity $x_n \equiv \frac{T}{2} \ln (Z_n^+/Z_n^-)$, where $Z^{\pm}_n$
are defined as the partition functions of a branch of generation $n$
with a fixed up (down) spin $s_0$ at the central site
\cite{BRU-84,SWI-94}, and is given by an implicit integral equation.
The fixed point equation is solved by numerical integration, and the
magnetization is computed for different values of $R$ using
Eq.~(\ref{magb}), for $T=0$ and $z=4$. In Fig.~\ref{figbethe} we show
a comparison between the magnetization in the GS and in the DS
(\cite{COL-02}).

%fig2

\begin{figure}[h]
\centerline{\psfig{file=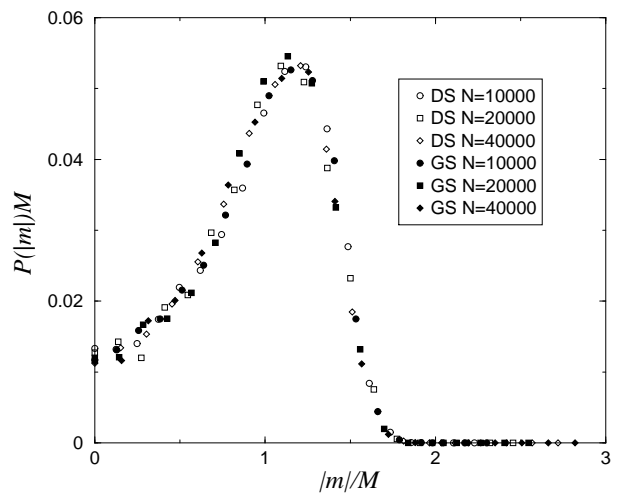,width=8cm}}
\caption{The distributions of the magnetization in the DS and the GS 
at their respective critical points on the Bethe lattice, obtained 
numerically for different lattice sizes $N$, can be all collapsed together.}
\label{figbethe2}
\end{figure}

As in $d=3$ simulations, the transitions occur at two different
locations (see the inset of Fig.~\ref{figbethe}), for $z=4$
$R_c^{(DS)}=1.781258...$ \cite{COL-02} and $R_c^{(GS)}\simeq 1.8375$, with
the mean-field exponent ($\beta=1/2$). When plotted against 
$(R-R_c)/R_c$ the two curves superimpose close to the critical point. 
This indicates that, though not required by universality, in the Bethe 
lattice $A_{GS}=A_{DS}$, as also found in $d=3$.  
To investigate possible
finite size scaling we have performed numerical simulations in the
Bethe lattice, following the method of Ref.~\cite{DHA-97}.  Collapsing
the order parameter curve as in $d=3$, using a scaling form similar to
Eq.~(\ref{eq:fss}), does not appear to be possible in the Bethe
lattice, because the scaling region is very narrow.  Thus to test
finite size scaling, we have computed the distribution of the 
magnetization $m$ at the respective critical point, $R_c^{(DS)}$ and
$R_c^{(GS)}$ for different lattice sizes $N$.  The distributions can
all be collapsed into the same curve (see Fig.~\ref{figbethe2}), using
the form $P(|m|)=f(|m|/M)/M$.

%fig3

To conclude we provide evidence about the universality of the RFIM
with Gaussian distributed disorder in and out of equilibrium.  The key
point is the identification of the correct order parameter for the
non--equilibrium transition.  This quantity, the remanent magnetization of the
DS, is the natural counterpart to the magnetization of the GS in the
equilibrium case, in particular at a zero external field. Our results
are based on a detailed numerical analysis in $d=3$ and on an exact
solution on the Bethe lattice.  It would be interesting to confirm
this conclusion by more complex measures, beside the order parameter,
such as the scaling of the domain wall stiffness \cite{MID-02}.
Regardless of the question of universality, the most intriguing point
from our analysis is that the two transitions appear at different
critical values of the disorder strength: the DS ferromagnetic phase
is the first to disappear as the disorder is increased
($R_c^{(DS)}<R_c^{(GS)}$).  The interpretation of the ordering is
simply that the GS is as correlated as a state can be in the RFIM, due
to the global optimization. Thus eg. as $R$ is decreased it is natural
that the FM correlations appear in the GS first.

Our results have important consequences on the use of the
demagnetization as an optimization tool: the difference in
the location of the critical points
implies that for $R_c^{(DS)} \leq R \leq R_c^{(GS)}$, the DS
(paramagnetic) is drastically different from the GS (ferromagnetic),
suggesting that in that region hysteretic optimization is likely to
fail. Moreover, to compare the GS and DS in a system, one
expects to achieve the closest resemblance in this regime
if the correlation length is the same; ie. due to the
difference of the critical points at two separate
values $R_1^{(DS)} <R_2^{(GS)}$, respectively -- or
as well two values of the effective coupling constant
$J$ in the Hamiltonian.

In addition to the ferromagnetic RFIM model, one
can speculate about other systems where two disorder induced phase 
transitions exist. Numerical simulations and analytical results have
shown that a disorder induced transition in the hysteresis loop can be
observed in the random bond Ising model \cite{VIV-94}, in the random
field O(N) model \cite{DAS-99}, in the random anisotropy model
\cite{VIV-01} and in the random Blume-Emery-Griffith model
\cite{VIV-94}. All these systems also show a transition in equilibrium
and it would be interesting to compare their DS and GS. Another
example would be the study of an interface in quenched disorder,
where many results are known for the roughness exponent in and out of
equilibrium (i.e. at the depinning threshold), and the results
typically differ \cite{NAT-00}. 
It would be interesting to measure the roughness of
an interface after a demagnetization cycle (i.e. after the field 
driving the interface is cycled with decreasing amplitude), and compare 
its properties with those of the ground state interface. 
Finally,  there is the issue of energetics of excitations
in the respective ensembles: the universality of exponents
and scaling functions would seem to imply that these also
scale similarly.

\end{document}